\documentclass[a4paper,10pt,twoside]{just}
\usepackage{multicol}
\usepackage{graphicx}
\usepackage{booktabs}
\usepackage{amssymb,bm,mathrsfs,bbm,amscd}
\usepackage[tbtags]{amsmath}
\usepackage{lastpage}

\begin{document}

\fancyhead[c]{\small  10th International Workshop on $e^+e^-$ collisions from $\phi$ to $\psi$ (PhiPsi15)}
 \fancyfoot[C]{\small PhiPsi15-\thepage}

\footnotetext[0]{Received 30 Nov. 2015}

\title{Measurements of $\pi^{0}\pi^{0}$ production and form factors for $f_{0}(980)$ and $f_{2}(1270)$ in single-tag two-photon process
 \thanks{Supported by National Natural Science Foundation of China}}

\author{%
      Qingnian Xu$^{1,2)}$\email{xuqingnian10@mails.ucas.ac.cn}
(for the Belle Collaboration)
}
\maketitle

\address{%
$^1$ Institute of High Energy Physics,  No.19B Yuquan Road, Beijing 100049, China \\
$^2$ University of Chinese Academy of Sciences,  No.19A Yuquan Road, Beijing 100049, China \\
}

\begin{abstract}
 We report a measurement of differential cross section of $\pi^{0}$ pair production in single-tag two-photon collisions.  
These results are obtained with 759$fb^{-1}$ of the data collected with the Belle detector at the KEKB asymmetric energy $e^{+}e^{-}$ collider. The cross section is
measured for $Q^{2}$ up to 30 GeV$^{2}$, where $Q^{2}$ is the negative of the invariant mass squared of the tagged virtual photon, in the kinematic range
0.5 GeV $<$ $W$ $<$ 2.1 GeV and $|$cos$\theta^{*}| <$ 1.0 for the total energy and pion scattering angle, respectively. The transition form factor of the
$f_{0}(980)$ and that of the $f_{2}(1270)$  with the helicity-0,-1, and -2 components separately are measured for the first time and are compared with theoretical
calculations.
\end{abstract}

\begin{keyword}
Single-tag, Two-photon, Cross section, Form factor.
\end{keyword}

\begin{pacs} 
12.38.Qk, 13.40.Gp, 14.40.Be
\end{pacs}

\begin{multicols}{2}

\section{Introduction}

The pseudo-scalar meson pair production via two-photon process $\gamma^{*}\gamma\rightarrow M\bar{M}$ in $e^{+}e^{-}$ collisions, with one virtual photon carrying large $Q^{2}$  and the other quasi-real photon, provides cleaning environments to probe the dynamics of hadronic interaction in the low energy region and to test QCD-based predictions. The cross sections of the $\gamma^{*}\gamma\rightarrow M\bar{M}$ production at LEP，B factory and charm experiments have been calculated with non-perturbative QCD approaches at large $Q^{2}$ and small $W$ \cite{QCD_apro}, where $Q^{2}$ is the negative of four-momentum squared of virtual photon and $W$ is the invariant mass of the meson pair. Clean determination of a gluon admixture in tensor mesons at large $Q^{2}$, by measuring the transition form factor (TFF) of the $f_{2}(1270)$, is suggested \cite{TFF_f1270}. A formalism for a model-independent evaluation of the subleading hadronic light-by-light (HLbL) contribution to the anomalous magnetic moment of the muon (g -2) is suggested recently \cite{dt_driven}. As a data-driven dispersive approach, it requires data as input and gives a more precise estimate of the HLbL scattering, that is supposed to dominate the theoretical error in the (g -2) calculation in future.  

Experimentally, this process can be studied with single-tag events, where one electron ($e^{\pm}$) is detected after emitting a virtual photon with a large $Q^{2}$  but the other ($e^{\mp}$) escaping in the forward direction. For those events, $Q^{2}$ is estimated with the negative of the measured four-momentum difference squared between the detected and the corresponding incoming electrons, and  $W$ is calculated from the measured invariant mass of the meson pair. In addition to analyze partial wave components, the valuable information on the $Q^{2}$ dependence of the TFF can be extracted from fitting the measured differential cross section of the $\gamma^{*}\gamma\rightarrow\pi^{0}\pi^{0}$ production.  In this analysis of  $e^{+}e^{-}\rightarrow e^{\pm}(e^{\mp})M\bar{M}$ events, the $Q^{2}$ value reaches up to 30 GeV$^{2}$, and $W$ mass is below 2.1 GeV.

\section{Belle detector and data sample}
  KEKB is an asymmetric $e^{+}e^{-}$ collider with beam energy 8.0 GeV for $e^{-}$ and 3.5 GeV for $e^{+}$ and the beam crossing  angle is $\pm11$ mr.
The Belle Detector~\cite{dete_1,dete_2}, is surrounding KEKB beams, covers the $\theta$ region from $17^{o}$ to $150^{o}$. The
Detector consist of a silicon vertex detector (SVD), a 50-layer central dirft chamber (CDC),an array of aerogel threshold Cherenkov counters (ACC), time-of-flight counters (TOF),
an electromagnetic calorimeter comprised of CsI(TI) crystals (ECL). These detectors are located inside a super-conducting solenoid coil that provide a 1.5T magnetic filed. An iron flux return located outside the coil is instrumented to detect $K_{l}$ mesons and identify muons.

We use a 759$fb^{-1}$ data sample recorded with the Belle detector \cite{dete_1,dete_2} at the KEKB asymmetric-energy $e^{+}e^{-}$ collier \cite{eecolider}. We combine data samples collected at several beam energies: at the $\Upsilon(4S)$ resonance, and 60 MeV below it (637$fb^{-1}$ in total); at the $\Upsilon(3S)$ resonance (3.2 $fb^{-1})$; and near the $\Upsilon(5S)$ resonance (119$fb^{-1}$). In this case combining the data, the slight dependence of the
two-photon cross section on beam energy is taken into account.

 We refer to events tagged by an $e^{+}$ or an $e^{-}$ as "p-tag"(positron-tag) or "e-tag"(electron-tag), respectively. We brief event selection criteria and show preliminary comparison between data and signal MC at first, and then present the results of the differential cross section and form factors.   Details in the analysis can be referenced to the Belle paper~\cite{orig_paper}.

\section{Event selection}
\subsection{Selection criteria}

 A pre-selection criterion is applied to select the signal events within the kinematical regions of $e^{+}e^{-}\rightarrow e(e)\pi^{0}\pi^{0}$ in which one electron escapes detection at small forward angles. We require exactly one charged track that satisfies $p_{t}$ $>$ 0.5 GeV/c in a required angular range, and no other tracks with $p_{t}$ $>$ 0.1 GeV/c in that range. Here, $p_{t}$ is the transverse momentum in the laboratory frame with respect to the positron beam axis. Energy sum of neutral clusters in the ECL is greater than 0.5 GeV. 

For electron ID, we require $E/p$ $>$ 0.8 for the candidate electron track. The absolute value of the momentum of the electron, corrected for photon radiation or bremsstrahlung, must be greater than 1.0 GeV/c. 

In order to reduce combination background in events with high multiplicity of neutral tracks, specific procedures for $\pi^{0}$ reconstruction are performed to suppress neutral background track, particularly fake photons with low energy \cite{orig_paper}. In addition to usual selection conditions for photon, as well as rejection of Bhabha events with a back-to-back configuration of an electron and $\pi^{0}$ in the $e^{+}e^{-}$ c.m. frame in which a track is not reconstructed, the following requirements for $\pi^{0}$ reconstruction must be satisfied. 

(1) The polar angle of the photons constrained to be at least one $\pi^{0}$ in the sensitive region of the ECL triggers. 

(2) In a second $\pi^{0}$ search, only one more $\pi^{0}$ is found among the $\pi^{0}$ candidates from the mass-constrained fit, and which does not share any photons with the first selected $\pi^{0}$.

(3) In case of more than two $\pi^{0}$ assigned, the one with the highest-energy photon is chosen. If still more than two combinations that share the highest-energy photon, the one in which the other photon in $\pi^{0}$ has the higher energy is chosen.

(4) The tagged electron have the correct charge sign (“right-sign”) with respect to the beam from which it originates in the $e^{+}e^{-}$ c.m. frame.

(5) A kinematical selection of 0.85 $<$ $E_{ratio}$ $<$ 1.1 is applied, where $E_{ratio}$ is defined as 
 $$E_{ratio}=E^{*measured}_{\pi^{0}\pi^{0}}/E^{*expected}_{\pi^{0}\pi^{0}},$$

where $E^{*measured}_{\pi^{0}\pi^{0}}$ is the $e^{+}e^{-}$ c.m. energy of the  system measured directly, $E^{*expected}_{\pi^{0}\pi^{0}}$ is energy of the $\pi^{0}\pi^{0}$ system expected by kinematics without radiation and is obtained by assigning the measured invariant mass to the $\pi^{0}\pi^{0}$ system.

Finally, the requirement for transverse momentum balance in the $e^{+}e^{-}$ c.m. frame is satisfied: $|\Sigma p_{t}|$ $<$ 0.2 GeV/c, where $|\Sigma p_{t}|$ = $|p^{*}_{t,e} + p^{*}_{t,\pi^{0}1} + p^{*}_{t,\pi^{0}2}|$.

\subsection{Kinematical variables}

There are four kinematical variables,  $Q^{2}$, $W$, $|$cos$\theta^{*}|$ and $\phi^{*}$. 
The angles $|$cos$\theta^{*}|$ and $\phi^{*}$ are defined in the $\gamma^{*}\gamma$ c.m.frame.
$Q^{2}$ is calculated using the measured four-momentum of the detected electron($p_{e}$) from
\begin{eqnarray}
\label{q2eqn}
Q^{2}_{rec} &=& -(p_{beam} - p_{e})^{2}\nonumber\\[1mm]
&=& 2E^{*}_{beam}E^{*}_{e}(1+q_{tag}cos\theta^{*}_{e}),
\end{eqnarray}
where $q_{tag}$ is the tagged electron charge, $p_{beam}$ is the nominal four-momentum of the beam particle with the same charge as the detected electron. 

\subsection{Signal MC and comparison of distribution for selected signal candidates}
The signal Monte Carlo (MC) generator TREPSBSS, based on the MC code in Ref.~\cite{treps_MC}, is developed to match the single-tag configuration and to implements the formulas of Ref.~\cite{sig_tag_MC}. The efficiency for single-tag two-photon events of $e^{+}e^{-}\rightarrow e(e)X$ and the two-photon luminosity function for $\gamma^{*}\gamma$ collisions are calculated with the TREPSBSS code.

We show some distributions for selected signal candidates in data and signal MC. Figs.~\ref{q2fig}, \ref{pi_energy} and \ref{pi_angle} show the $Q^{2}$, pion energy and pion angle distributions for data sample and corresponding signal MC in three $W$ regions for e- and p-tag. 
Since backgrounds are not subtracted in the experimental data, the comparability between the data and signal MC is reasonably satisfied in the kinematical region concerned.
\begin{center}
\includegraphics[width=7cm]{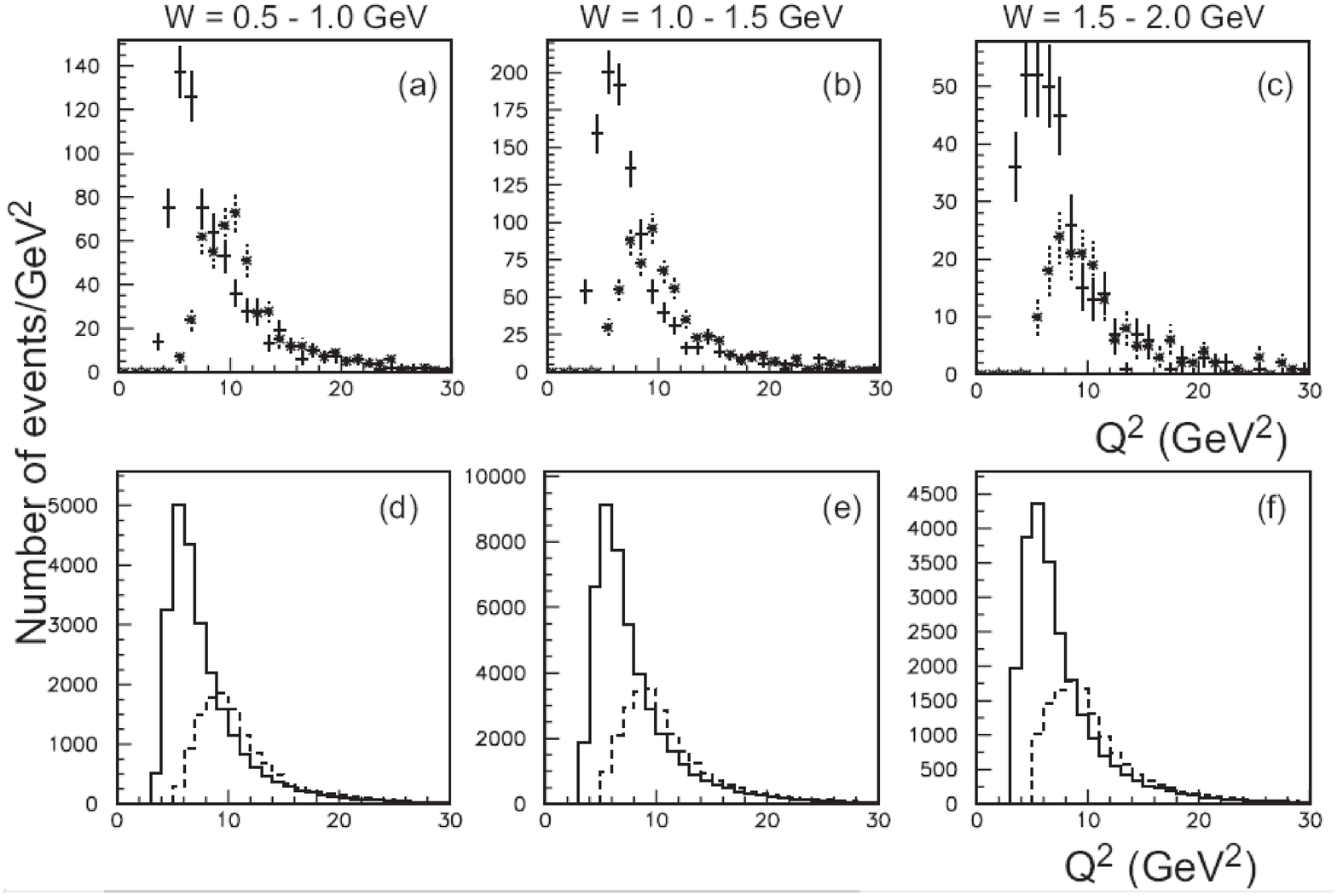}
\figcaption{\label{q2fig} Top three plots(a,b,c) are the $Q^{2}$ distributions for data in three $W$ regions. The cross and asterisk plot are for e-tag and p-tag samples, respectively. Bellow three plots(d,e,f) are the corresponding signal MC distribution, the solid and dashed histograms are for e-tag and p-tag samples,respectively}
\end{center}

\begin{center}
\includegraphics[width=7cm]{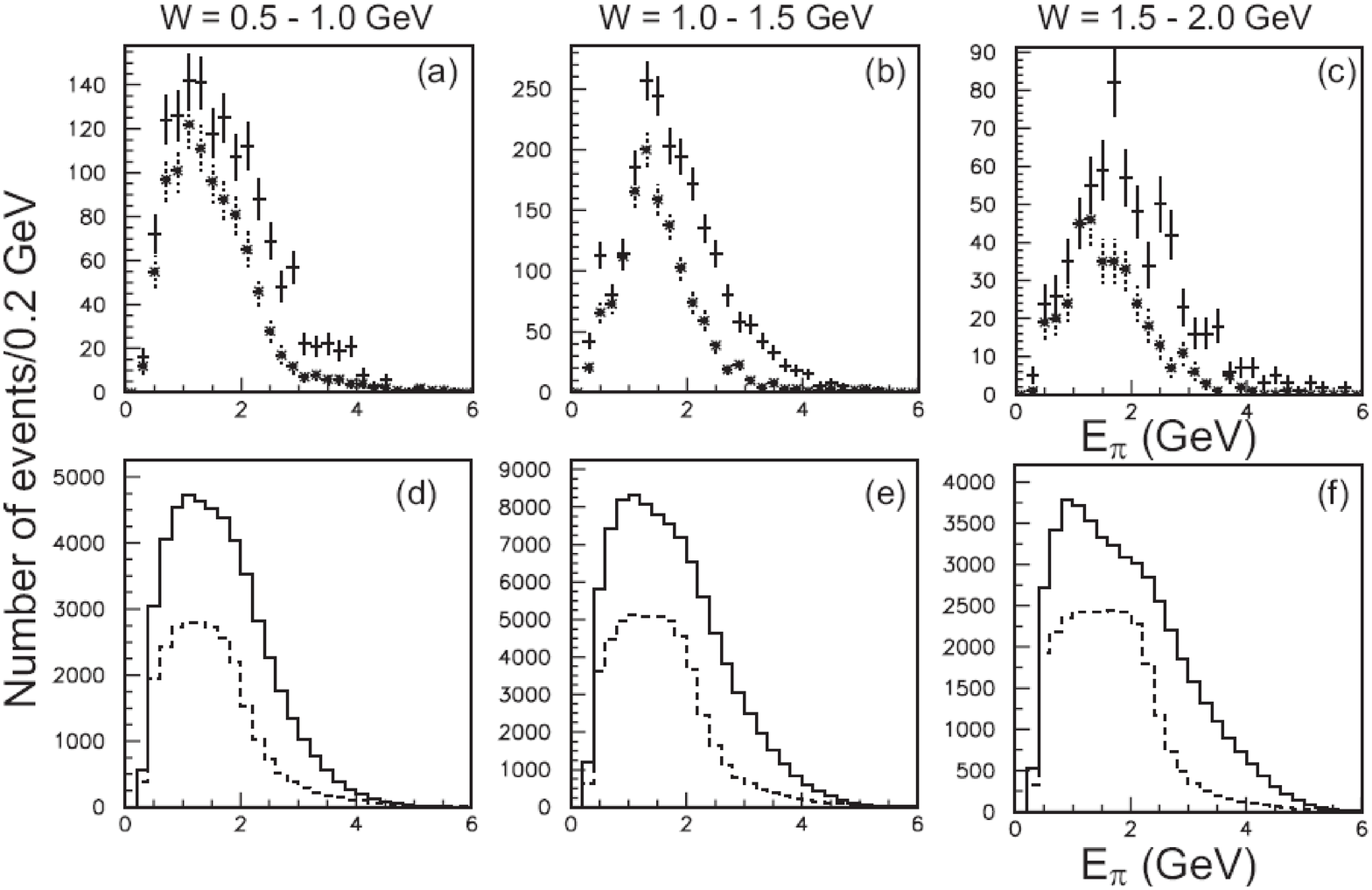}
\figcaption{\label{pi_energy} Top three plots(a,b,c) are the laboratory energy  distributions for two $\pi^{0}$ in three $W$ regions. Bellow three plots(d,e,f) are the corresponding signal MC distribution. The legend and comments are the same as those in Fig.~\ref{q2fig}.}
\end{center}

\begin{center}
\includegraphics[width=7cm]{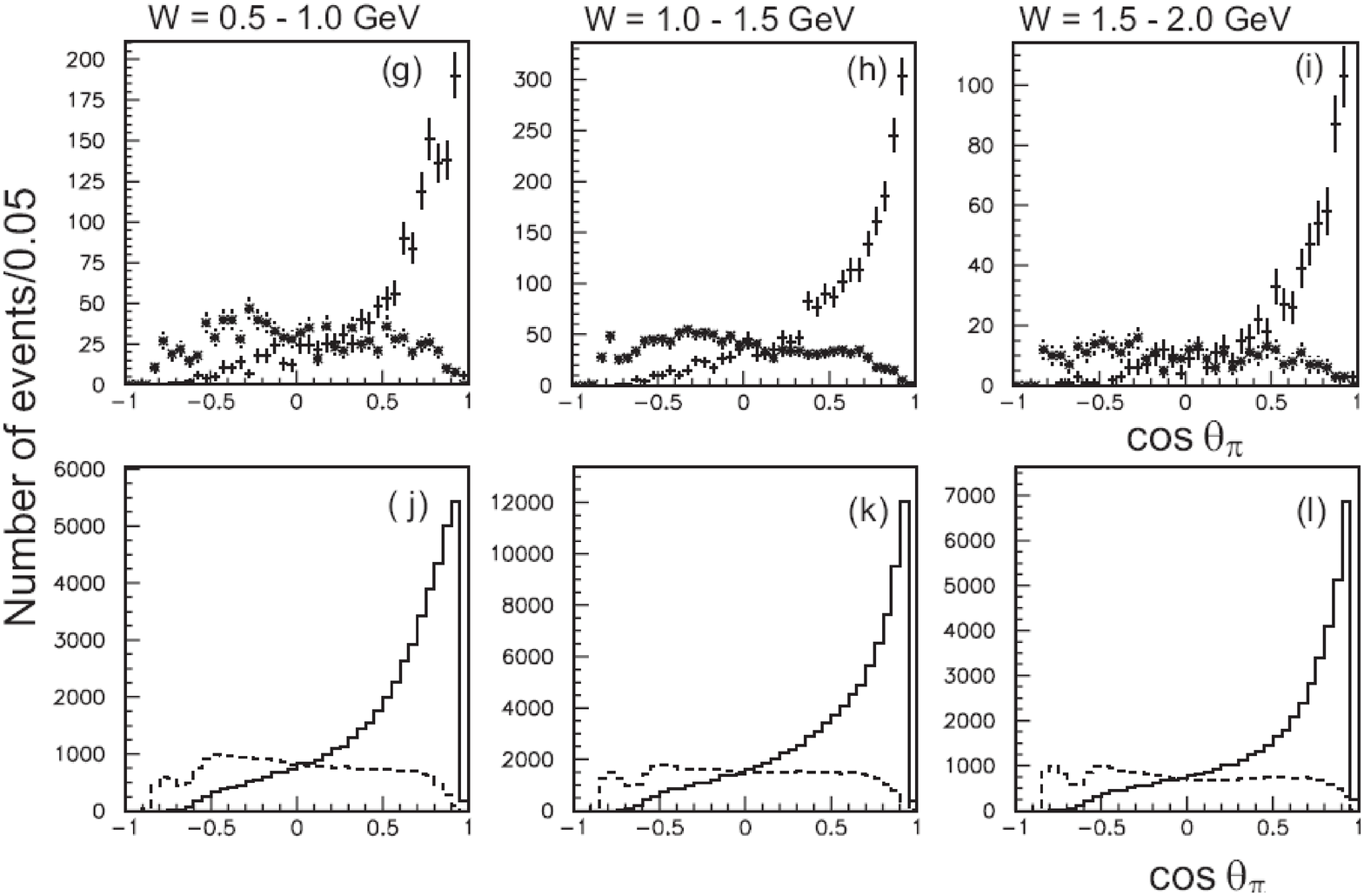}
\figcaption{\label{pi_angle} Top three plots(g,h,i) are the laboratory angle  distributions for two $\pi^{0}$ in three $W$ regions. Bellow three plots(j,k,l) are the corresponding signal MC distribution. The legend and comments are the same as those in Fig.~\ref{q2fig}.}
\end{center}

\subsection{Background estimation}

Various background sources are carefully studied and their contributions are estimated \cite{orig_paper}. 
The background contributions from single pion production via $e^{+}e^{-}\rightarrow e(e)\pi^{0}$ and the three-$\pi^{0}$ production process are small, and thus their effects are estimated and included in the systematic uncertainty. 
Some part of the backgrounds from the virtual Compton process (with fake or noise photon hits) and from the non-exclusive processes (from the tail part of the low $E_{ratio}$ activity) has been corrected by subtracting the estimated quantity.
The background contribution from  $\pi^{0}\gamma$ production via $e^{+}e^{-}\rightarrow e(e)\pi^{0}\gamma$ is estimated and subtracted from observed signal yield.

\section{Measurement of the differential cross section}

 The $e^{+}e^{-}$-based cross section is defined separately for the p-tag and -e-tag as follows:
\begin{eqnarray}
\label{eebasdcr}
&&(\frac{d^{3}\sigma_{ee}}{dWd|cos\theta^{*}|dQ^{2}})_{x-tag} = \nonumber\\[1mm]
&&\frac{Y_{x-tag}(W,|cos\theta^{*}|,Q^{2})}{\epsilon'_{x-tag}(W,|cos\theta^{*}|,Q^{2})\Delta W \Delta|cos\theta^{*}|\Delta Q^{2}\int LdtB^{2}},
\end{eqnarray}
where the yield Y and the uncorrected efficiency obtained by the signal MC 
$\epsilon'$ are separately evaluated for p-tag and e-tag, for a consistency check. $\int Ldt$ is the integrated luminosity of 759 $fb^{-1}$ and $B^{2}$ = 0.9766 is the square of the decay branching fraction $B(\pi^{0})\rightarrow\gamma\gamma$.

After confirming the consistency between the p- and e-tag measurements and taking the difference of the beam
energies in evaluating $\epsilon'$ into account, , we combine the yield and the efficiency using the formula for both measurements,
\begin{eqnarray}
\label{cmbep}
&&\frac{d^{3}\sigma_{ee}}{dWd|cos\theta^{*}|dQ^{2}} = \nonumber\\[1mm]
&&\frac{Y(W,|cos\theta^{*}|,Q^{2})（1-b(W,|cos\theta^{*}|,Q^{2}))}{\epsilon'(W,|cos\theta^{*}|,Q^{2})\Delta W \Delta|cos\theta^{*}|\Delta Q^{2}\int LdtB^{2}},
\end{eqnarray}
where $Y = Y_{p-tag} + Y_{e-tag}$, $\epsilon' = (\epsilon'_{p-tag} + 
\epsilon'_{e-tag})/2$ and $b$ is the background fraction combined for p- and
e-tags, which is subtracted here.

Dividing by the single-tag two-photon luminosity function $d^{2}L_{\gamma^{*}\gamma}/dWdQ^{2}$, the $e^{+}e^{-}$-based differential cross section is converted to that $\gamma^{*}\gamma$-based cross section:
\begin{eqnarray}
\label{ggbcr}
&&\frac{d\sigma_{\gamma^{*}\gamma}}{d|cos\theta^{*}|} = \nonumber\\[1mm]
&&\frac{d^{3}\sigma_{ee}}{dWd|cos\theta^{*}|dQ^{2}}\frac{f}{2\frac{d^{2}L_{\gamma^{*}\gamma}}{dWdQ^{2}}(1+\delta)(\epsilon/\epsilon')\epsilon'}.
\end{eqnarray}
The factors $\delta$, $\epsilon$, and $f$ correspond to the radiative
correction, efficiency corrected for the $\phi^{*}$ dependence of the differential cross section, and the unfolding effect that accounts for migrations between the different $Q^{2}$ bins, respectively.

 Figs.~\ref{eebcr_cos} and \ref{eebcr_w} compare the $e^{+}e^{-}$-based cross section measured separately for the p- and e-tags.
The detector acceptance is much different between e- and p-tag, and the cross-section measured for the two tags are consistent within statistical errors. This provide a validation check for trigger, detector acceptance, and selection conditions.

Fig.~\ref{ggbcr_w} shows the $\gamma^{*}\gamma$-based cross section as a function of $W$ in nine $Q^{2}$ bins. Peaks corresponding to the $f_{2}(1270)$ and $f_{0}(980)$ are evident.

\begin{center}
\includegraphics[width=7cm]{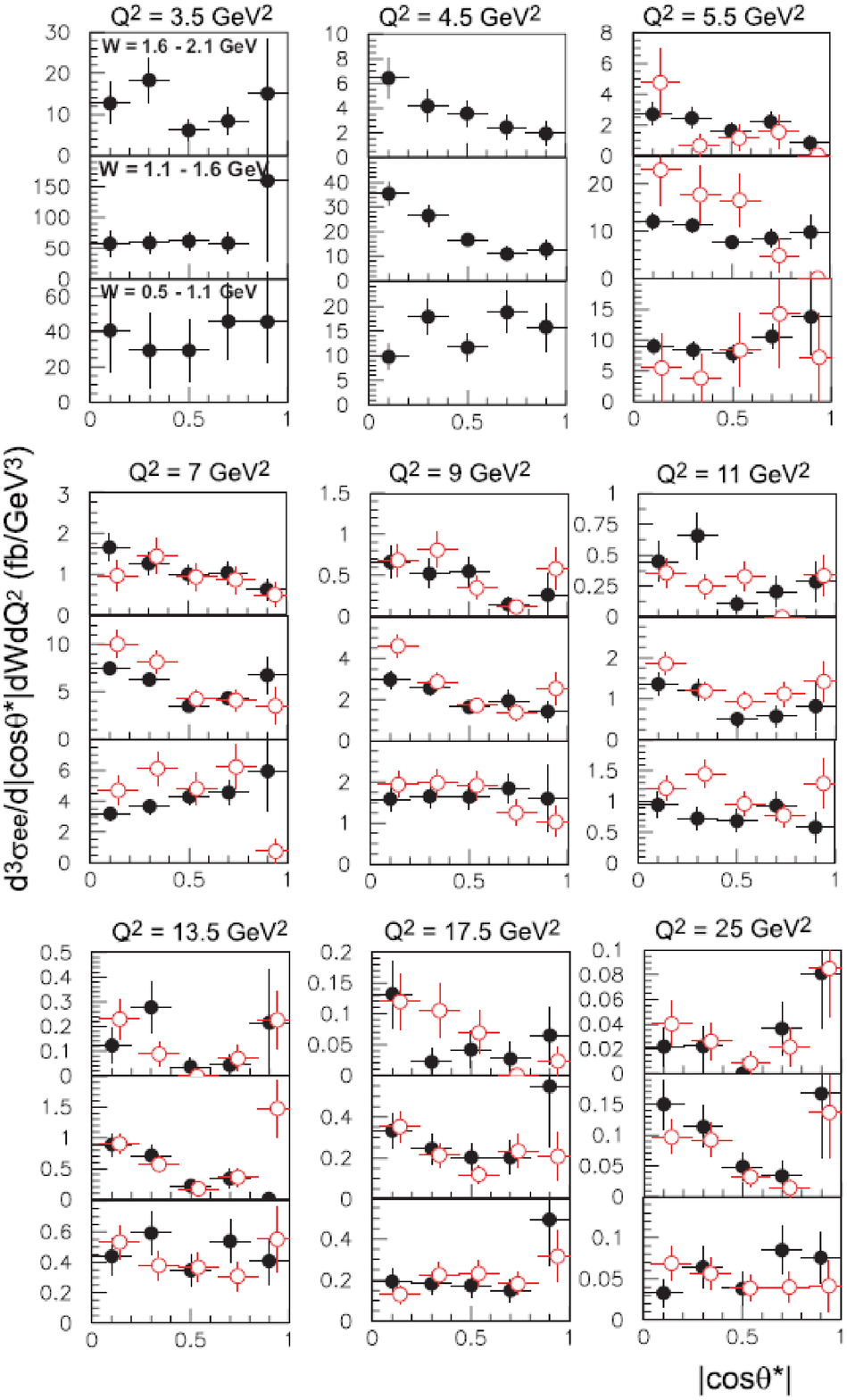}
\figcaption{\label{eebcr_cos}  The $|$cos$\theta^{*}|$ dependence of the $e^{+}e^{-}$-based cross section in each $Q^{2}$ bin in three selected  $W$ regions, 0.5 $<$ $W$ $<$ 1.1 GeV, 1.1 $<$ $W$ $<$ 1.6 GeV, 1.6 $<$ $W$ $<$ 2.1 GeV from bottom to top. The closed circles (open circles) are for the e-tag (p-tag) measurements.}
\end{center}

\begin{center}
\includegraphics[width=6cm]{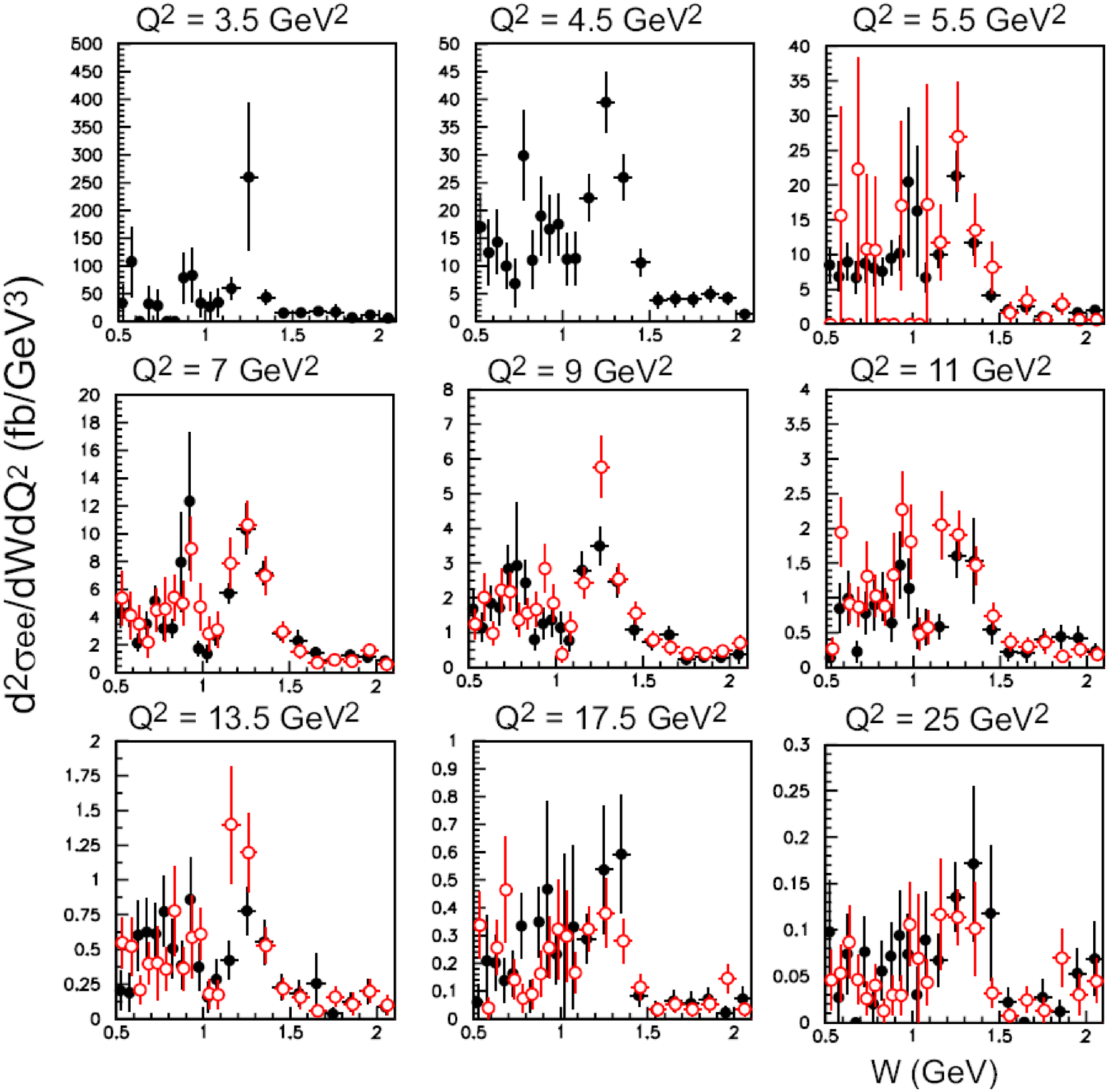}
\figcaption{\label{eebcr_w} The $W$ dependence of the $e^{+}e^{-}$-based cross section in each $Q^{2}$ bin, with $|$cos$\theta^{*}|$ in range 0 to 1, is integrated. The closed circles (open circles) are for the e-tag (p-tag) measurements.}
\end{center}

\begin{center}
\includegraphics[width=7cm]{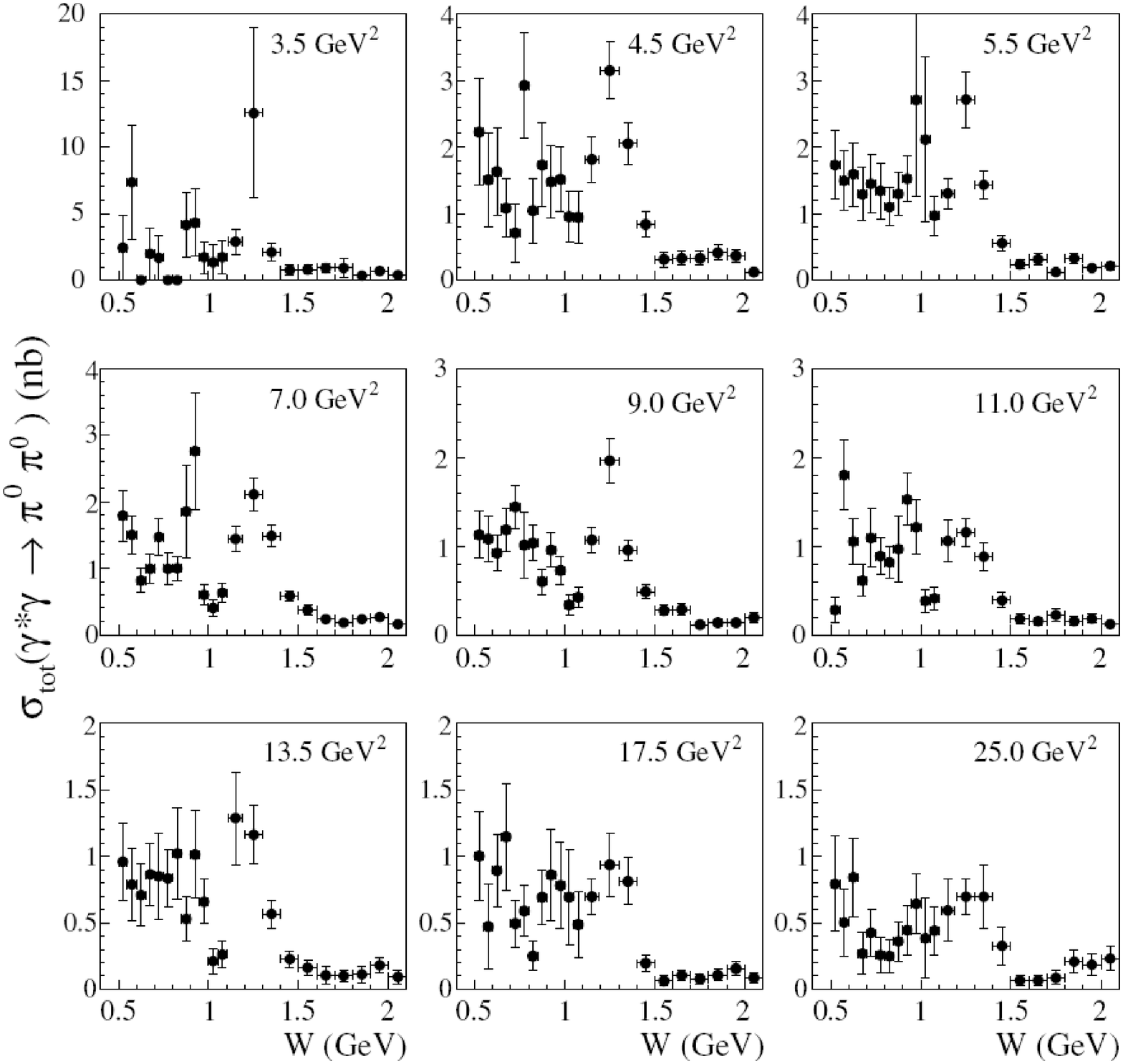}
\figcaption{\label{ggbcr_w}  Integrated cross section for $\gamma^{*}\gamma\rightarrow\pi^{0}\pi^{0}$ in nine $Q^{2}$ bins in GeV$^{2}$ indicated in each panel.}
\end{center}

\section{Measurement of transition form factors}

 In this section, we extract the $Q^{2}$ dependence of the TFF of the $f_{0}(980)$ and those of the helicity-0, -1, and -2 components of the  $f_{2}(1270)$.

The differential cross section for the process  $e^{+}e^{-}\rightarrow e^{+}e^{-}\pi\pi$  is given by Ref.~\cite{dcr_fun}, and its formalism can be found in Ref.~\cite{orig_paper}. As  partial waves with $J$ above 2 in the region of $W$ $\leq$ 1.5 GeV are negligibly small, only even angular-momentum partial waves, $S$ and $D$, contribute.

The $Q^{2}$ dependent TFFs, $F_{f2}(Q^{2})$ of the $f_{2}(1270)$ together with its helicity-0, -1, and -2 components and $F_{f0}(Q^{2})$ of the $f_{0}(980)$, are extracted from fitting the differential cross section of the data in the energy region 0.7 GeV $<$ $W$ $<$ 1.5 GeV. The fits are done by parameterizing ($S$, $D_{0}$, $D_{1}$ and $D_{2}$) and ($S$, $D_{1}$ and $D_{2}$) waves, respectively.

$S$ and $D_{i}$ ($i$ = 0, 1, 2) are parametrized as follows:

\begin{eqnarray}
\label{pwa_par_sdi}
S &=& A_{f_{0}(980)}e^{i\phi f_{0}} + B_{S}e^{i\phi_{BS}}, \nonumber\\[1mm]
D_{i} &=& \sqrt{r_{i}(Q^{2})}A_{f_{2}(1270)}e^{i\phi_{f_{2}Di}} + B_{Di}e^{i\phi_{BDi}}.
\end{eqnarray}
where $A_{f_{0}(980)}$ and $A_{f_{2}(1270)}$ are the amplitudes of the $f_{0}(980)$ and $f_{2}(1270)$, respectively; $r_{i}(Q^{2})$ is the fraction of the $f_{2}(1270)$-contribution in the $D_{i}$ wave with the constraints $r_{0}$ + $r_{1}$ + $r_{2}$ = 1  and $r_{i}$ $>$ 0; $B_{S}$ and $B_{Di}$ are non-resonant "background" amplitudes for the $S$ and $D_{i}$ waves; $\phi_{BS}$, $\phi_{BDi}$, $\phi_{f0}$ and $\phi_{f2Di}$ are the phases of background amplitudes, $S$ and $D_{i}$, the $f_{0}(980)$ and $f_{2}(1270)$ in the $D_{i}$ waves, respectively. The phases are assumed to be independent of $Q^{2}$.

The parametrizations of $Q^{2}$  dependence TFFs for  $f_{0}(980)$ and $f_{2}(1270)$ give in Refs.\cite{par_A_1, par_A_2}, for the $f_{0}(980)$  the parametrization is adopted as:

\begin{eqnarray}
\label{pwr_A_f0980}
A_{f_{0}(980)} = F_{f0}(Q^{2})\sqrt{1+\frac{Q^{2}}{M^{2}_{f0}}}\frac{\sqrt{8\pi\beta_{\pi}}}{W}\frac{gf_{0\gamma\gamma}gf_{0\pi\pi}}{16\sqrt{3}\pi}\frac{1}{D_{f_{0}}}
\end{eqnarray}

The parametrization for the f2(1270) is given in Ref.~\cite{par_A_f21270}. The relativistic Breit-Wigner resonance amplitude $A_{R}(W)$ is given by 
\begin{eqnarray}
\label{pwr_A_f0980}
A^{J}_{R}(W) &=& F_{R}(Q^{2})\sqrt{1+\frac{Q^{2}}{M^{2}_{R}}}\sqrt{\frac{8\pi(2J+1)m_{R}}{W}} \nonumber\\[1mm]
&&\times\frac{\sqrt{\Gamma_{tot}(W)\Gamma(W)B(\pi^{0}\pi^{0})}}{m^{2}_{R} - W^{2} - im_{R}\Gamma_{tot}(W)}
\end{eqnarray}

We can extract information on partial waves for three out of the four ($S$, $D_{0}$, $D_{1}$ and $D_{2}$) waves only, because of a limitation in the analysis with the $\phi^{*}$-integrated cross section. To partially overcome this limitation, we first fit the $\phi^{*}$-dependent (but $Q^{2}$-integrated) differential cross section to obtain information on the fractions of the $f_{2}(1270)$ in the $D_{0}$, $D_{1}$ and $D_{2}$ waves. Then, this information is used in the fit of the $\phi^{*}$-integrated cross section.

Figure~\ref{cr_f1270_hel0}, \ref{cr_f1270_hel1} and \ref{cr_f1270_hel2} show the $Q^{2}$ dependence of TFF for $f_{2}(1270)$ with its  helicity-0, -1, -2 components. For helicity-2 component the measured  TFF agrees well with the prediction by Ref.~\cite{the_f1270_hel2_1} and with one of two predictions by Ref.~\cite{the_f1270_hel2_2}. For helicity-0, -1 components the measured TFFs are about a factor of 1.5 - 2 smaller than the prediction by Ref.~\cite{the_f1270_hel2_1}.

Figure~\ref{cr_f980} shows the obtained $Q^{2}$ dependence of
the TFF of the $f_{0}(980)$. The result is agree well with the prediction by  Ref.~\cite{the_f1270_hel2_1} for $Q^{2}\leq$ 10 GeV$^{2}$, but has less steeper $Q^{2}$ for $Q^{2}$ $>$ 10 GeV$^{2}$.

\begin{center}
\includegraphics[width=7cm]{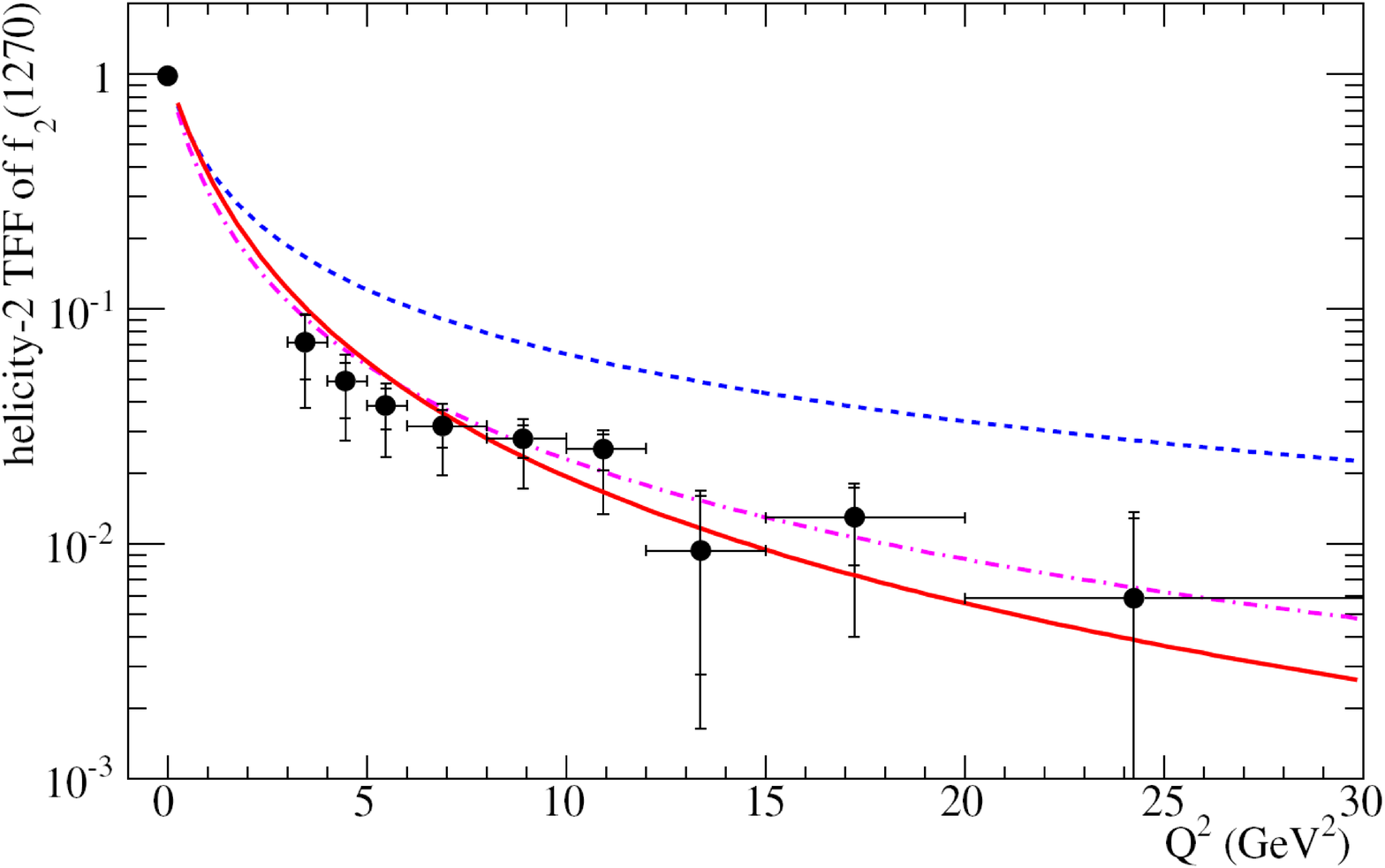}
\figcaption{\label{cr_f1270_hel2}  The measured helicity-2 TFF of the $f_{2}(1270)$ as a function of $Q^{2}$. Short  (long) vertical bars indicate statistical (statistical and systematic combined) errors. The solid line shows the predicted $Q^{2}$  Ref.~\cite{the_f1270_hel2_1} and those by Ref.~\cite{the_f1270_hel2_2} dashed line and dot-dashed line. }
\end{center}

\begin{center}
\includegraphics[width=7cm]{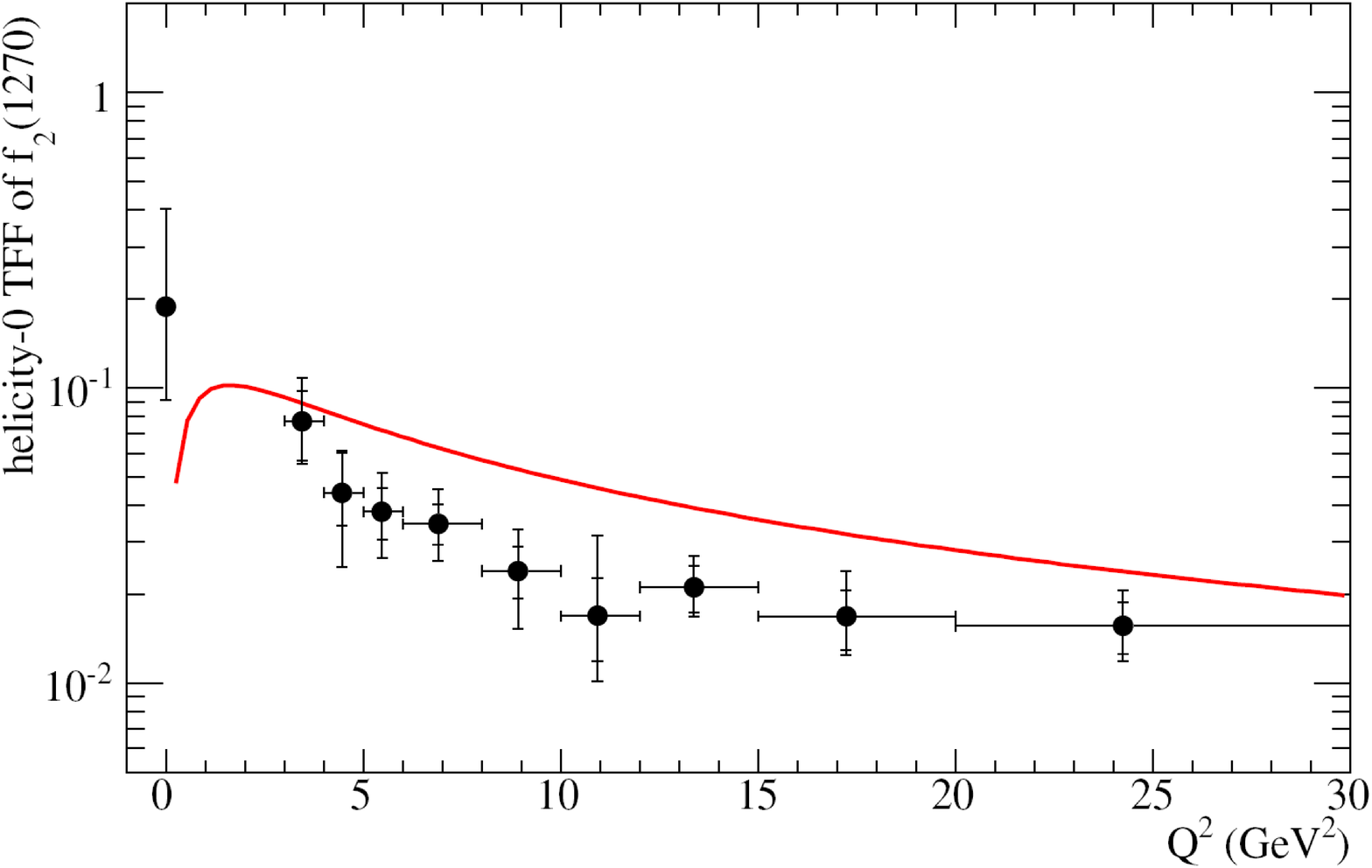}
\figcaption{\label{cr_f1270_hel0} The measured helicity-0 TFF of the $f_{2}(1270)$
as a function of $Q^{2}$. Short (long) vertical bars indicate statistical (statistical and systematic combined) errors. The solid line shows the predicted $Q^{2}$ dependence by Ref.~\cite{the_f1270_hel2_1}.}
\end{center}

\begin{center}
\includegraphics[width=7cm]{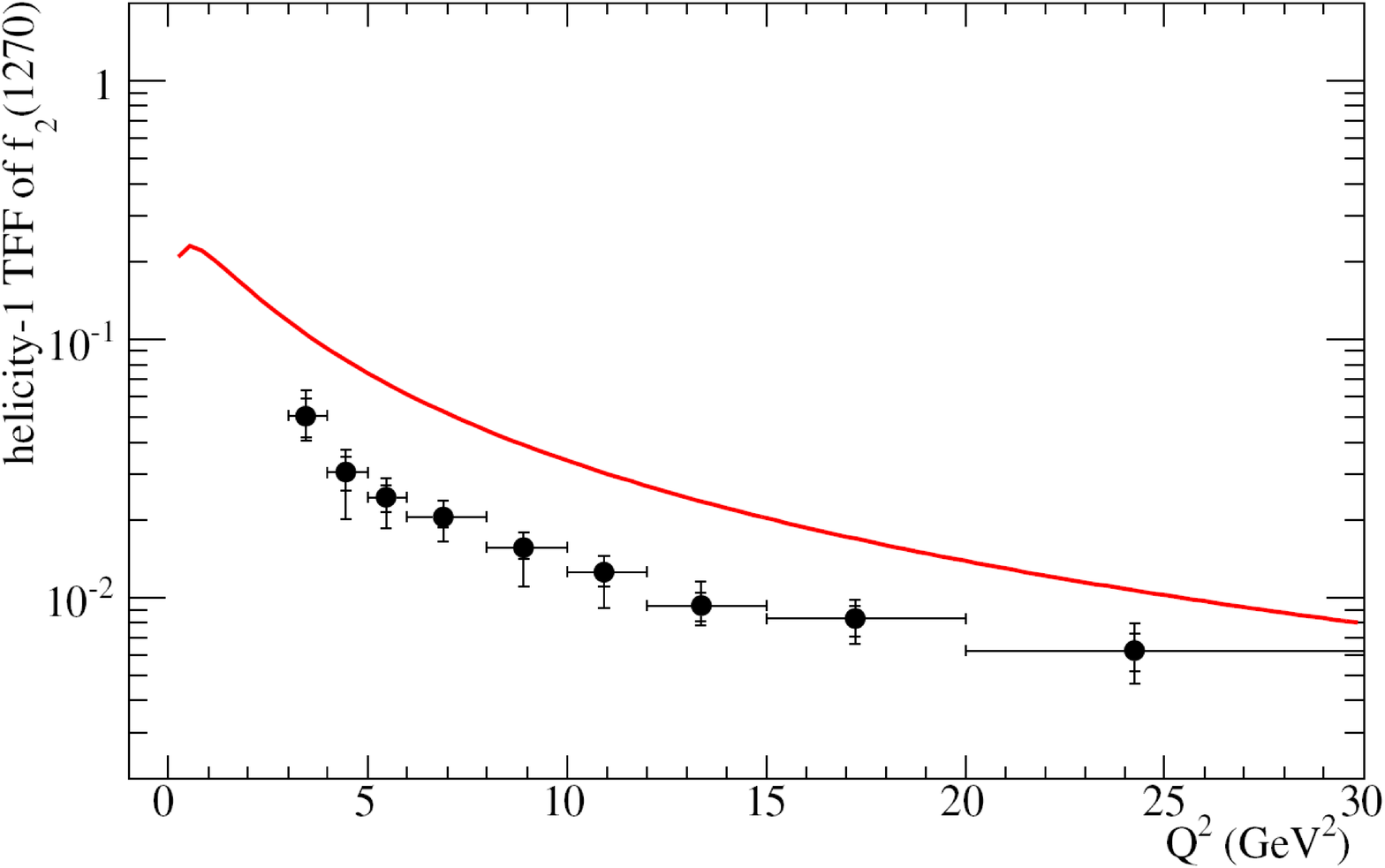}
\figcaption{\label{cr_f1270_hel1} The measured helicity-1 TFF of the $f_{2}(1270)$
as a function of $Q^{2}$. Short (long) vertical bars indicate statistical (statistical and systematic combined) errors.The solid line shows the predicted $Q^{2}$ dependence by Ref.~\cite{the_f1270_hel2_1}.}
\end{center}

\begin{center}
\includegraphics[width=7cm]{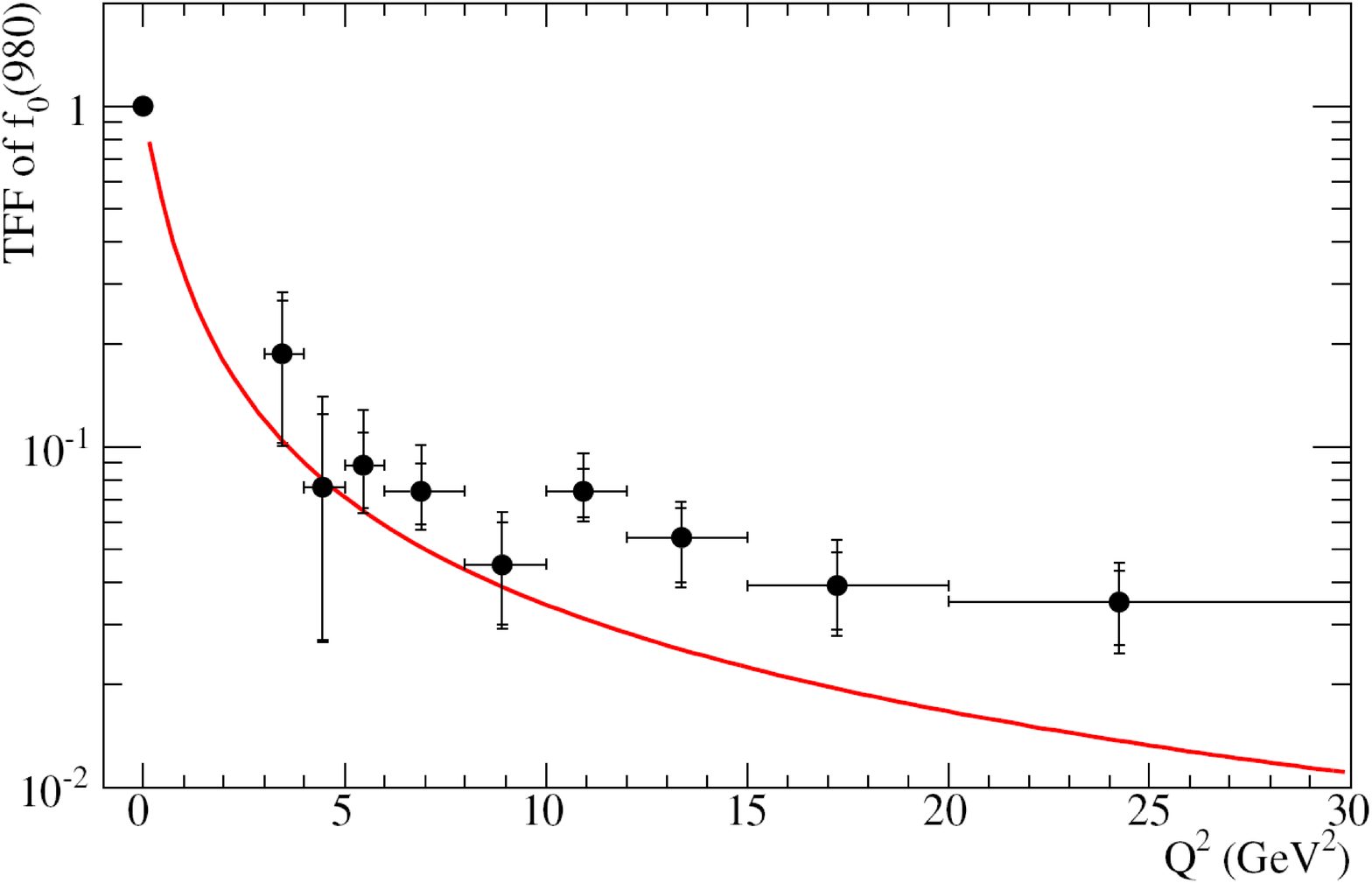}
\figcaption{\label{cr_f980} The measured $Q^{2}$ dependence of the TFF of the $f_{0}(980)$. Short (long) vertical bars indicate statistical (statistical and systematic combined) errors. The solid line shows the prediction for a scalar particle given  Ref.~\cite{the_f1270_hel2_1}.}
\end{center}

\section{Conclusion}

The differential cross section of  $\gamma^{*}\gamma\rightarrow\pi^{0}\pi^{0}$  production in single-tag mode with $Q^{2}$ up to 30 GeV$^{2}$ is measured for the first time, based on a data sample of 759 $fb^{-1}$ collected with the Belle detector \cite{dete_1, dete_2} at the KEKB asymmetric-energy $e^{+}e^{-}$ collider \cite{eecolider}. The kinematical variables, $W$ and $|$cos$\theta^{*}|$ in the $\gamma^{*}\gamma$ c.m  system, covers a range of 0.5 GeV to 2.1 GeV and less than 1.0, respectively.

Significant signals for $f_{0}(980)$ and $f_{2}(1270)$ are clearly seen in the angle-integrated cross section distribution of the $\gamma^{*}\gamma\rightarrow\pi^{0}\pi^{0}$ production in different $Q^{2}$ regions. Analysing the $\phi^{*}$-dependent differential cross section with the $S$ and $D$ waves included, the results show that the contribution of the helicity-0 component of the $f_{2}(1270)$ is large, while that of its helicity-1 component is small but non-zero.

The differential cross section is fitted by parameterizing partial-wave amplitudes. The transition form factors (TTF) of the $f_{2}(1270)$ and $f_{0}(980)$ are measured for $Q^{2}$ up to 30 GeV$^{2}$ and compared with theoretical predictions. The resulting helicity-2 TFF of the $f_{2}(1270)$ agrees well with either the prediction based on a heavy quark approximation        \cite{the_f1270_hel2_1}, or one of the two predictions from the formulated sum rules \cite{the_f1270_hel2_2}. The helicity-0 and -1 TFF of the $f_{2}(1270)$ are about a factor of 1.5 - 2 smaller than the prediction of Ref.~\cite{the_f1270_hel2_1}. The $Q^{2}$ dependence of the $f_{0}(980)$ TFF agrees fairly well with the prediction of Ref.~\cite{the_f1270_hel2_1} for  $Q^{2} \leq$ 10 GeV$^{2}$ but has less steeper $Q^{2}$ dependence for $Q^{2}$ $>$ 10 GeV$^{2}$.

\end{multicols}

\begin{multicols}{2}

\end{multicols}

\clearpage

\end{document}